# *COBE* Constraints On Baryon Isocurvature Models [†]

Wayne Hu,[1] Emory F. Bunn[1], Naoshi Sugiyama[1,2]

[1]*Department of Astronomy and Physics
and Center for Particle Astrophysics*

[2]*Department of Physics, Faculty of Science
The University of Tokyo, Tokyo 113, Japan*

We consider Bayesian constraints on standard isocurvature baryon models from the slope and normalization of the anisotropy power spectrum detected by the *COBE* DMR experiment in their two year maps. In conjunction with either the amplitude of matter fluctuations $\sigma_8$ *or* its slope, all open models are ruled out at greater than 95% confidence, whereas cosmological constant $\Lambda$ dominated models are constrained to be highly ionized. By including the *COBE* FIRAS 95% confidence upper limit on spectral distortion, we further reduce the available parameter space for $\Lambda$ models by excluding these highly ionized models. These constraints define a single class of barely viable standard models which makes definite and testable predictions for degree scale anisotropies and large scale structure.

*Subject Headings: cosmic microwave background - large scale structure of universe*

---

*Obtaining life is a matter of timeliness.*
*Losing it is a matter of compliance.*
*Repose in timeliness and dwell in compliance*
*Then sorrow and joy can never enter.*

*–Chuang-tzu*

## 1. Introduction

The original baryon isocurvature scenario for structure formation (Peebles 1987a,b) presents a simple and attractive alternative to the standard cold dark matter (CDM) cosmogony. It simultaneously satisfies observations which require a low density universe $\Omega_0 \simeq 0.2 - 0.3$ (*e.g.* Dekel *et al.* 1993), forms structure without the aid of hypothetical dark matter, and can alter light element nucleosynthesis sufficiently to make an $\Omega_0 = \Omega_b$ baryonic universe acceptable (Gnedin & Ostriker 1992). Moreover recent measurements of a large Hubble constant $H_0 = 100h$ km s$^{-1}$ Mpc$^{-1}$, $h = 0.80 \pm 0.17$ (Freedman *et al.* 1994) would be easier to accommodate in such a low density universe.

Unfortunately, when normalized to the *COBE* DMR detection (Smoot *et al.* 1992), the open universe manifestations of this model appear to be inconsistent with several observations of cosmic microwave background anisotropies at degree (Chiba, Sugiyama, & Suto 1994; Hu & Sugiyama 1994, hereafter HS94), and arcminute scales (Efstathiou & Bond 1987; Hu, Scott, & Silk 1994). The model generically suffers from excess small scale power. However, given the present uncertain status of CMB anisotropy detections at degree to arcminute scales (see *e.g.* Wilkinson 1994), it is perhaps premature to rule out models on these grounds.

The excess small scale power required in this scenario also appears as a steep slope in the large angle anisotropy spectrum (Sugiyama & Silk 1994). This prediction conflicts with the flat spectrum measured by the *COBE* DMR experiment (Gorski *et al.* 1994; Bunn, Scott, & White 1995). In this *Letter*, we quantify this constraint on the standard isocurvature baryon model by employing the techniques of Bunn & Sugiyama (1995) to analyze the two year *COBE* DMR maps. Unlike previous treatments (Chiba, Sugiyama, & Suto; HS94), we also use the full information in the *COBE* sky maps to determine the normalization as opposed to merely the rms fluctuation at 10 degrees. This causes a 10% boost in the amplitude of fluctuations in open models. However we further extend prior treatments by considering flat low $\Omega_0$, cosmological constant $\Lambda$ models whose predictions are somewhat more in accord with observations. The boost in amplitude can be up to 30% in these models. The corresponding change in the matter fluctuation amplitude $\sigma_8$ is relevant for simulations of large scale structure formation. Finally, employing spectral distortion constraints from the *COBE* FIRAS experiment (Mather *et al.* 1994), we nearly close the parameter space available to these baryon isocurvature models. For the small class of models remaining, we present the predictions for degree scale anisotropies and large scale structure.

## 2. General Features

In the standard baryon isocurvature model, the universe consists of photons, baryons, and three families of massless neutrinos only. Initial entropy perturbation, *i.e.* fluctuations in the baryon-photon and baryon-neutrino number densities, are assumed to take the form of a pure power law in $\tilde{k}$, $|S(\tilde{k})|^2 \propto \tilde{k}^n$ where the wavenumber $\tilde{k}$ is related to the eigenvalue of the Laplacian $k$ as $\tilde{k}^2 = k^2 + K$, with $K = -H_0^2(1 - \Omega_0 - \Omega_\Lambda)$ as the curvature (Wilson 1983). Here $\Omega_\Lambda$ is the fraction of the critical density contributed by the cosmological constant. The $\Lambda$ dominated models which we consider here are flat for simplicity, *i.e.* $K = 0$. In this case, $\tilde{k} = k$ and represents an ordinary Fourier mode of the perturbation. This limit is also appropriate for large scale structure measurements which are affected by perturbations on scales that are well under the curvature radius.

Since there is no *ab initio* mechanism for generating the entropy perturbations, the index $n$ is fixed by measurements of large scale structure today. Isocurvature perturbations evolve such that below the photon diffusion scale, the initial entropy fluctuations become the density perturbations that seed large scale structure. This implies that the observational constraints of an $P(k) \simeq k^{-1}$ power spectrum at large scale structure scales (*e.g.* Peacock & Dodds 1994) implies an $n \simeq -1$ initial power law in the model. Numerical simulations which take into account non-linear modifications of this picture confirm this result (Suginohara & Suto 1992). At the largest scales however, isocurvature conditions prevent the formation of density perturbations leading to a steep $P(\tilde{k}) \propto (\tilde{k}^2 - 4K)^2 \tilde{k}^n$, *i.e.* an $n + 4$ power spectrum below the curvature scale. This sharply rising spectrum of fluctuations on scales relevant for CMB anisotropies is particularly dangerous when normalized at the large scales by the *COBE* DMR measurement.

It may thus seem that the model can be ruled out by merely considering the implied amplitude of the matter power spectrum at the $8h^{-1}$Mpc scale. In an unbiased scenario of galaxy formation, which is expected in these baryon only models (Cen, Ostriker, & Peebles 1994), observations require $\sigma_8 \simeq 1$. However, the baryon isocurvature model has an additional degree of freedom to save it. Since Silk damping (Silk 1968) does not destroy entropy fluctuations, the large amount of small scale power in the model allows for collapse of objects immediately following recombination. This could lead to sufficient energy input to reionize the universe (Peebles 1987). Because Compton drag prevents the growth of structure, the ionization history can be tuned to provide the right ratio of matter to temperature fluctuations. Following Gnedin & Ostriker (1992), we assume that a fraction $x_e$ of the electrons were reionized at $z \simeq 800$. For complications due to a multi-staged ionization history and compact baryonic object formation, see HS94.

Reionization also leads to significant and observable consequences for the CMB. Large primary fluctuations from the acoustic oscillation phase (see *e.g.* Hu & Sugiyama 1995) are exponentially damped with optical depth below the horizon at the new last scattering surface. Secondary anisotropies are generated due to Doppler shifts off of moving electrons at last scattering. These are damped under the thickness of the last scattering surface due to cancellation of redshifts and blueshifts as the photon travels across many wavelengths of the perturbation. Thus it is almost always the case that higher ionization implies smaller anisotropies under the angle that the horizon subtends at last scattering. We plot the anisotropies in a $\Lambda$ model as a function of $x_e$ in Fig. 1, where the rms anisotropy is related to $C_\ell$ via $\langle |\Delta T/T|^2 \rangle = \sum (2\ell+1) C_\ell / 4\pi$, with $\ell$ as the multipole number of the spherical harmonic decomposition of anisotropies on the sky. Open universe examples are displayed in HS94.

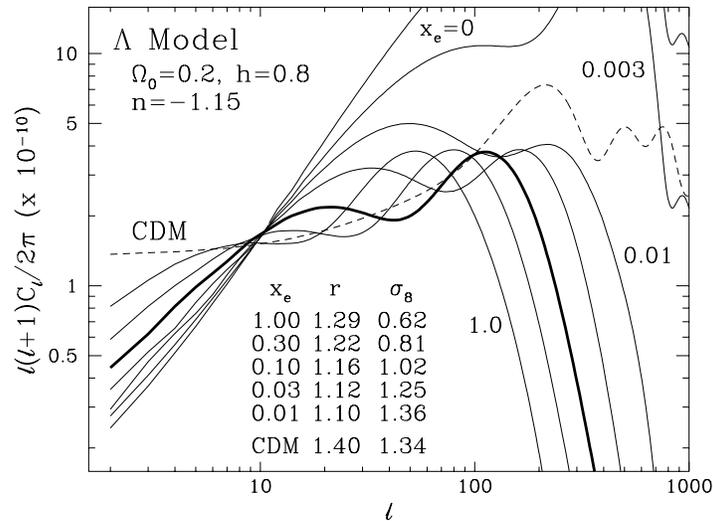

**Figure 1.** *COBE* normalized anisotropies in the $\Lambda$ model as a function of ionization. As the ionization level increases from $x_e = 0$ to 1 as listed in the figure, the damping reaches to larger angles making the *COBE* slope shallower. Open universe models suffer less from this effect at large angles due to geodesic deviation. Fluctuations are also regenerated on the new last scattering surface. With high enough ionization, this can once again steepen the *COBE* slope. The *COBE* normalization also sets the level of matter fluctuations $\sigma_8$ for a fixed thermal history. The ratio $r$ of amplitudes between the more complete likelihood analysis used here and the $10°$ rms normalization is shown. The most promising model which currently escapes constraints from the *COBE* slope, normalization, and spectral distortion measurements is shown (thick line, $x_e = 0.1$). This should be compared with the standard CDM ($h = 0.5$, $\Omega_b = 0.05$, $Q_{\mathcal{L}} = 19.9\mu\mathrm{K}$) model (dashed line). Measurements in the range $\ell \simeq 20 - 200$ can further help to distinguish the models.

One exception to this damping rule is the second order Doppler contributions from the Vishniac effect (Ostriker & Vishniac 1986; Vishniac 1987) which is not included in Fig. 1. This effect is uncovered at arcminute scales, where other first order effects have suffered severe thickness damping, and is extremely sensitive to the amplitude of the matter perturbations. It thus is only important for highly ionized, late last scattering scenarios (Hu, Scott, & Silk 1994; HS94).

Finally ionization also implies that the electrons have been heated to a temperature above that of the CMB. This implies that Compton scattering will lead to spectral distortions in the CMB as photons are upscattered in frequency by the electrons. The distortion is described by the Compton-$y$ parameter defined as $y = \int d\tau\, k(T_e - T)/m_e c^2$ where $T_e$ and $T$ are the electron and CMB temperatures respectively and $\tau$ is the optical depth to Compton scattering. Explicitly $d\tau = (x_e n_e \sigma_T c)dt$, where $x_e$ is the ionization fraction, $n_e$ is the electron number density, and $\sigma_T$ is the Thomson cross section. Thus the greater the level of ionization, the larger the spectral distortion in the CMB.

# 3. Model Constraints

When extended to large scales, the steep initial spectrum required by large scale structure conflicts with the flat anisotropy spectrum measured by the *COBE* DMR experiment. Without reionization, the predicted *COBE* DMR slope is approximately $n_{\rm eff} \simeq 2$ (Sugiyama & Silk 1994) compared with observational constraints of $n_{\rm eff} = 1.3^{+0.24}_{-0.37}$ (with quadrupole, Bunn, Scott, & White 1995). The fact that $n_{\rm eff}$ is only weakly dependent on $n$ and is somewhat shallower than one might expect from the $n+4$ behavior of the matter power spectrum is discussed in Hu & Sugiyama (1995).

Reionization tends to suppress small angle anisotropies and can mitigate a steep initial spectrum. However, if the ionization is too great, secondary anisotropies generated on the last scattering surface will counter this effect (see Fig. 1). Furthermore, reionization has no effect for angles much larger than that subtended by the horizon at last scattering. At the *COBE* scale, open models will thus be less affected by reionization than $\Lambda$ models, since geodesic deviation carries the same physical scale at last scattering to a much smaller angle on the sky today. Lesser effects can be attributed to raising the baryon content through $\Omega_b h^2$ which delays last scattering and increases the physical scale of the horizon. However even for flat models, projection effects due to the distance to the last scattering surface depend strongly on $\Omega_0$ and counters the $\Omega_b$ dependence in these $\Omega_0 = \Omega_b$ baryonic models. Furthermore, the late integrated Sachs-Wolfe effect (Sachs-Wolfe 1967; Hu & Sugiyama 1995) boosts the low order multipoles slightly as $\Omega_0$ decreases. In the range of interest, decreasing $\Omega_0$ leads to a shallower *COBE* slope. High $x_e$, high $h$, low $\Omega_0$, $\Lambda$ models therefore offer the best prospects of bringing down the *COBE* slope.

Employing the two year *COBE* DMR sky maps and the analysis methods developed by one of us (EB), we can place an upper limit on the primordial index $n$ for open and $\Lambda$ models fixed by $\Omega_0$, $h$, and $x_e$. We expand the two-year DMR data in a set of basis functions which are optimized to have the maximum rejection power for incorrect models (Bunn & Sugiyama 1995; Bunn, Scott, & White 1995). We then use the 400 most significant terms in this expansion to compute the likelihood functions for a variety of models. To set limits on $n$ and the normalization $Q$, the rms quadrupole, we assume a prior distribution which is uniform for all $Q$ and $n \leq 0$. Spectra with $n > 0$ are unphysical due to non-linear effects which regenerate an $n=0, P(k) \propto k^4$ large scale tail to the fluctuations (Zel'dovich 1965; Peebles 1980). The constraint in the crucial $n \simeq -1$ regime is not sensitive to the details of this cutoff. Shown in Fig. 2 are the 95% confidence upper limits imposed on $n$ by integrating over the normalization $Q$ to form the marginal likelihood in $n$. As expected, all open models with $n \simeq -1$ are ruled out regardless of ionization fraction, whereas highly ionized $\Lambda$ models remain acceptable.

With the maximum likelihood value for the normalization $Q_\mathcal{L}$ of the model, we predict the amplitude of matter fluctuations $\sigma_8$. The likelihood value for the normalization tends to boost the amplitude over the *COBE* DMR 10° rms normalization value of $30\mu$K (Bennett *et al.* 1994) by a factor $r \equiv Q_\mathcal{L}/Q_{10°} \simeq 1.1$ for open models and low ionization $\Lambda$ models. The difference is more significant in highly ionized $\Lambda$ models due to the damping of the anisotropy spectrum. The boost is on the order $r \simeq 1.3$ for a fully ionized $\Lambda$ model (see Fig. 1). This effect appears also in the CDM model with a greater magnitude in fact. The effect of the low quadrupole in the data on the 10° measure (Bunn, Scott, & White 1995) artificially supresses the amplitude. In all cases, the likelihood analysis provides the better normalization by including the full data set and minimizing the effects of cosmic variance. In Fig. 2, we thus plot the value of $\sigma_8$ corresponding to this normalization as a function of ionization history and spectral index. The suppression of fluctuation growth in a highly ionized universe must be compensated by a steeper spectral index $n$. Notice that even ignoring limits on the large scale structure slope, all open models which satisfy the *COBE* slope are ruled out.

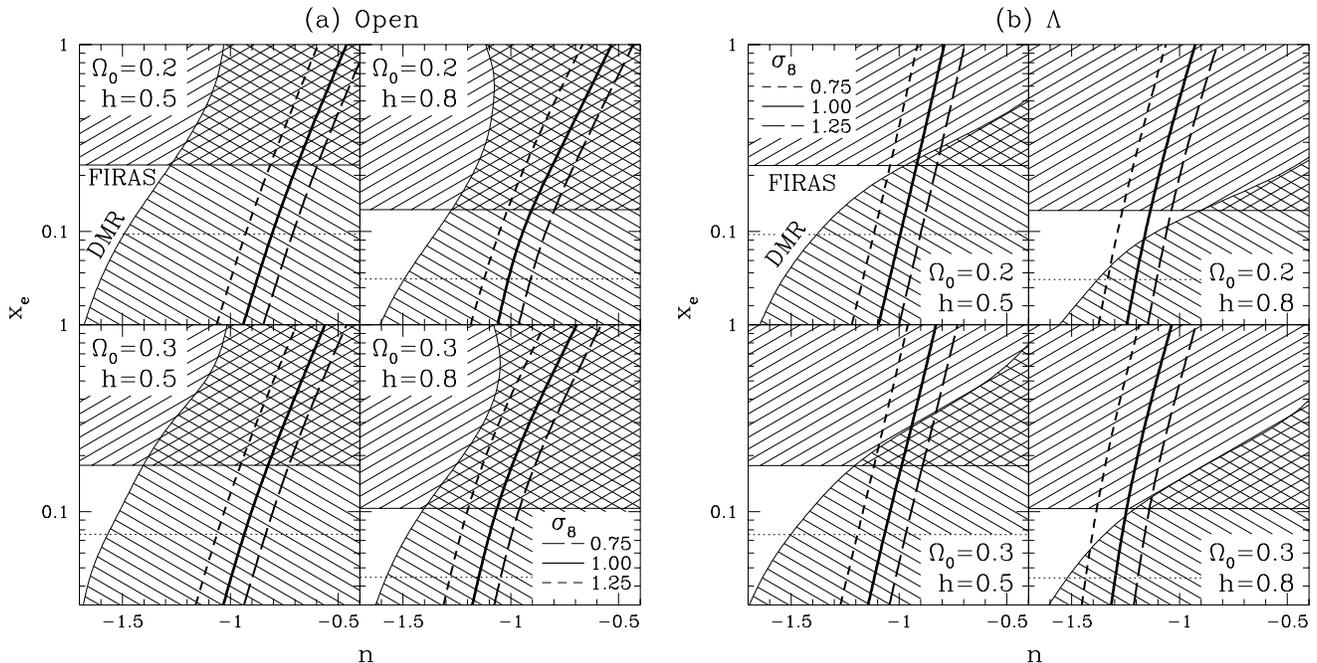

**Figure 2.** Constraints on the primordial spectral index $n$ and ionization fraction $x_e$. The *COBE* DMR slope imposes a 95% upper confidence limit on $n$ which is weakened as the ionization fraction increases, due to damping of the primary fluctuations, until a turning point at which fluctuations are significantly regenerated by the Doppler effect on the new last scattering surface. The *COBE* FIRAS constraint on spectral distortions through the Compton-$y$ parameter sets an upper limit on the ionization fraction. Here a conservative $T_e = 5000$K is assumed. The more realistic $T_e = 10000$K is also shown (dotted lines). The *COBE* DMR normalization also sets the level of matter fluctuations at the 8 $h^{-1}$Mpc scale $\sigma_8$. (a) No open model simultaneously satisfies all the observational constraints. (b) For $\Lambda$ models, a small region of parameter space is open for high $h$, low $\Omega_0$ models. The full anisotropy spectrum for the most promising model $\Omega_0 = 0.2$, $h = 0.8$, $n = -1.15$ and $x_e = 0.1$ is displayed in Fig. 1 and the matter power spectrum in Fig. 3. Even this model is ruled out with the more realistic $T_e$.

Even though highly ionized $\Lambda$ models can survive constraints on the *COBE* slope and the large scale structure normalization, they run into difficulties with the low upper limit on spectral distortions imposed by the *COBE* FIRAS experiment, $y < 2.5 \times 10^{-5}$ (95% CL). If the intergalactic medium is collisionally ionized, the electron temperature must be $T_e \gtrsim 10,000$K (see *e.g.* Gnedin & Ostriker 1992). To be conservative, we take $T_e = 5000$K. The corresponding limit from the Compton-$y$ parameter may be avoided by more exotic ionization schemes which attempt to inject as little energy as possible into the electrons (*e.g.* neutrinos decaying via a 13.6 eV photon). However since we generically expect at least a few eV excess energy above the ionization threshold, 5000K is a reasonable minimal estimate of the electron temperature. Calculations by Tegmark & Silk (1994) of photo-ionized models, which include feedback from the CMB through Compton cooling, support this conclusion. With this constraint, even $\Lambda$ models fall from favor. Only high $h$ models have a small window of parameter space open, which in fact closes if a more realistic $T_e = 10000$K is assumed. For these early ionized scenarios, this constraint largely obviates the need to impose limits from the Vishniac effect (HS94).

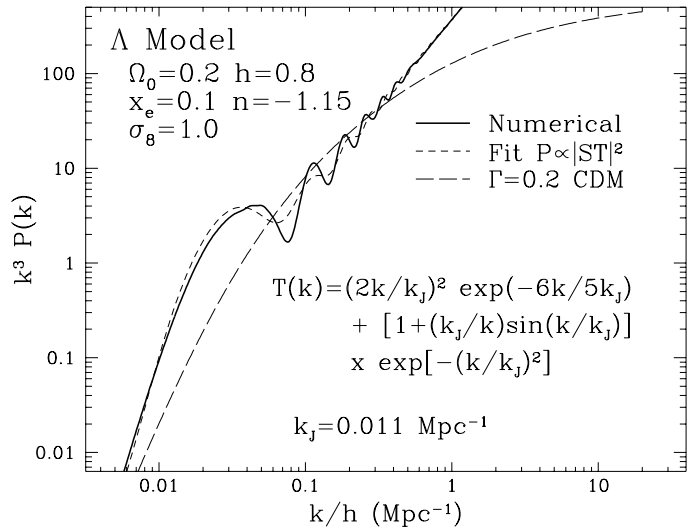

**Figure 3.** Power spectrum $k^3 P(k)$ for an allowed $\Lambda$ model ($\Omega_0 = 0.2$, $h = 0.8$, $x_e = 0.1$, $n = -1.15$). For comparison, a parameterized $\Gamma = 0.2$ CDM model (Efstathiou, Bond, & White 1992) is shown normalized to $\sigma_8 = 1$ which is known to fit the shape of the large scale structure data at $10^{-2} \lesssim k/h \lesssim 1 \text{Mpc}^{-1}$. To facilitate more detailed comparisons, we have also provided a simple fitting formula involving the isocurvature transfer function $P(k) \propto |T(k)S(k)|^2$ and the maximal Jeans scale $k_J$ which is motivated by the perturbation analysis of Hu & Sugiyama (1995).

Finally, it is worthwhile to mention that once the observational situation at degree scales settles down, one can at the very least impose a lower limit on the ionization fraction (see Fig. 1 and HS94). Significant reionization ($x_e \gtrsim 0.01$) is necessary in these models to avoid large degree and arcminute scale fluctuations in the CMB. Moreover, with the rapidly increasing number of experiments, the sample variance associated with the measurements (Scott, Srednicki, & White 1994) will decrease to the point where all baryon isocurvature models can be distinguished from the CDM model. In Fig. 1, we have plotted the predictions for the most promising baryon isocurvature model (thick solid line) in comparison to the CDM model (dashed line). The shape to the rise of the prominent peak around $\ell \simeq 100 - 200$ and the fall off thereafter may be used to help distinguish the models. In Fig. 3, we plot the matter power spectrum for the same model and include a simple fitting formula which may facilitate the comparison with large scale structure measurements. As a simple comparison, we also show a $\Gamma = 0.2$ CDM model which is known to fit the slope of the observed spectrum well (Efstathiou, Bond, & White 1992). These models may consequently run into problems with the large and small scale regimes of large scale structure measurements as well as indications of a smooth power spectrum (Peacock & Dodds 1994). With these additional constraints, one may hope to close off the already small window of parameter space available to the model.

# 4. Discussion

Standard baryon isocurvature models generically run into conflict with CMB observations even at *COBE* scales. Open models are ruled out by a combination of the *COBE* spectral slope and implications of the normalization for the matter power spectrum. Whereas these two considerations leave a large window of acceptable $\Lambda$ models, the inclusion of the *COBE* FIRAS constraint on spectral distortions even in a relatively conservative fashion is sufficient to drastically reduce the available parameter space such that a tuning of the ionization history, $\Omega_0$ and $h$ must be involved.

With the current state of affairs in which none of the simplest models for structure formation fare well in comparison with *all* the observations of the CMB and large scale structure, it is perhaps unwise to dismiss this scenario as entirely unviable. The general idea of isocurvature seeded fluctuations may of course be saved by introducing more free parameters.

The original model employs two simplifying assumptions: a power law initial spectrum and a constant ionization fraction after reionization. Since open models run into difficulties by predicting a steep *COBE* slope, the former assumption must be dropped to save them. In fact, for the open models there is some reason to believe that the spectrum may possess non-trivial structure at the curvature scale (Lyth & Stewart 1990; Ratra & Peebles 1994; Bucher, Goldhaber, & Turok 1994). Note however that unlike the open adiabatic case, power law behavior in gravitational potential fluctuations is equivalent to power law behavior in the entropy fluctuation (Hu & Sugiyama 1995), which serves to eliminate a potential ambiguity of the open model. Moreover, tentative indications of a break to a steep rise in anisotropies at degree scales (Scott & White 1994) may require further adjustment of the initial spectrum even below the curvature scale in the open model.

On the other hand, more complicated ionization histories and compact baryonic object formation can be employed to help design a more favorable $\Lambda$, but not open, model. The ionization history can fixed such that the relative normalization of the matter and radiation yields $\sigma_8 = 1$ (see HS94). However, since even the maximally damped open models violate the *COBE slope* constraint if $n \simeq -1$, no tuning of thermal histories alone can save the open baryon isocurvature scenario. For $\Lambda$ models, thermal history effects can also be employed to escape the *COBE* FIRAS constraints without giving up the damping benefits of a highly ionized model. This is because spectral distortions are a function of the total optical depth, whereas the damping of anisotropies is determined at last scattering where the optical depth equals unity. Thus late ionized scenarios may be more favorable. Unfortunately, the large amount of small scale power may make delayed reionization impossible.

More radical solutions have also been proposed. Peebles (1994) suggests the addition of cold dark matter or defects, which may also provide sufficient freedom to save the model. Small admixtures of adiabatic fluctuations may also be added. Yet with these *ad hoc* patches on the model, the appeal of the baryon isocurvature scenario is greatly reduced.

# Acknowledgments

We would like to thank D. Scott, J. Silk, and M. White for useful discussions. W.H. was supported by the NSF and N.S. by the JSPS.